# A memory-free spatial additive mixed modeling for big spatial data

Daisuke Murakami[1,*], Daniel A. Griffith[2]

[1]Department of Data Science, Institute of Statistical Mathematics,

10-3 Midori-cho, Tachikawa, Tokyo, 190-8562, Japan

Email: dmuraka@ism.ac.jp

[2]School of Economic, Political and Policy Science, The University of Texas, Dallas,

800 W Campbell Rd, Richardson, TX, 75080, USA

Email: dagriffith@utdallas.edu

* Corresponding author

Abstract

This study develops a spatial additive mixed modeling (AMM) approach estimating spatial and non-spatial effects from large samples, such as millions of observations. Although fast AMM approaches are already well-established, they are restrictive in that they assume a known spatial dependence structure. To overcome this limitation, this study develops a fast AMM with the estimation of spatial structure in residuals and regression coefficients together with non-spatial effects. We rely on a Moran coefficient-based approach to estimate the spatial structure. The proposed approach pre-compresses large matrices whose size grows with respect to the sample size $N$ before the model estimation; thus, the computational complexity for the estimation is independent of the sample size. Furthermore, the pre-compression is done through a block-wise procedure that makes the memory consumption independent of $N$. Eventually, the spatial AMM is memory-free and fast even for millions of observations. The developed approach is compared to alternatives through Monte Carlo simulation experiments. The result confirms the estimation accuracy of the spatially varying coefficients and group coefficients, and computational efficiency of the developed approach. Finally, we apply our approach to an income analysis using United States (US) data in 2015.

## 1. Introduction

Regression problems with thousands to millions of observations are currently common (Wood et al. 2015). The same holds for environmental science, econometrics, ecology, and other fields employing spatial data (i.e., data with location information). In fact, together with the development of sensing and positioning technologies, the availability of spatial data continues to increase. Spatial data are published through portal sites such as Google Earth Engine and OpenStreetMap. The development of regression approaches to handle big spatial data has become an urgent task.

Regression for spatial data has been studied for decades. Therein, spatial dependence, which refers to a stronger dependence between nearby things than distant things, is considered as the most basic property of spatial data (Anselin 2010). The Gaussian process (GP or kriging; see Cressie 1993) and other processes describing spatial dependence have been developed in geostatistics (Gelfand et al. 2010), spatial econometrics (LeSage and Pace 2009), and other fields. In regression analysis, ignorance of residual spatial dependence can lead to underestimation of coefficient standard errors and an increase in the risk of a Type I error (see LeSage and Pace 2009). Residual spatial dependence has been modeled using a spatially dependent process to avoid these problems (e.g., Seya and Tsutsumi 2008; 2009).

A spatial process has been assumed behind not only residuals but also other model elements. For example, the Bayesian spatially varying coefficient (B-SVC) modeling of Gelfand et al. (2003) is a representative approach to estimate spatially varying coefficients (SVCs) by assuming GPs behind regression coefficients. The Moran eigenvector-based SVC modeling of Murakami et al. (2017), which we introduce here later, is a faster alternative to the B-SVC modeling approach.

Beyond spatial dependence, consideration of non-spatial effects is also important.

Linear/non-linear, group, temporal, and many other effects are possible; it is required to correctly identify these effects behind data. Especially, Hox (1998), Goldstein (2011), among others, have suggested the importance of considering group effects whose ignorance can lead to erroneous inference (see Section 2.4). Following them, we focus on group effects later. Fortunately, the additive mixed model (AMM) framework that accounts for a wide variety of effects, including group effects, has been well-established in applied statistics (e.g., Hodges 2016; Wood 2017).

Based on the preceding observations, AMM estimating spatial and non-spatial effects behind thousands to millions of observations will be useful. Unfortunately, GP-based spatial modeling on a two-dimensional space requires a computational complexity of $O(N^3)$ and a memory storage of $O(N^2)$, where $N$ is the sample size; GP is not suitable for large samples. To lighten the computational burden, low-rank and sparse GPs have been developed in geostatistics. The former includes the predictive process (Banerjee et al. 2008), fixed rank kriging (Cressie and Johanesson 2008), and multiresolution approximations (e.g., Katzfuss 2017; Griffith and Chun 2019), whereas the latter includes a GP with covariance tapering (Furrer et al., 2006), the nearest-neighbor A GP (Datta et al. 2016), a Gaussian Markov random fields (GMRF)-based GP approximation (Lindgrn et al. 2016), and a $l_1$-penalized maximum likelihood approach (Krock, 2019), is related to the graphical lasso (Friedman et al., 2008). Yet, most geostatistical studies assume univariate GP without non-spatial effects (e.g., non-linear effects and group effects from covariates). It is unclear how to extend these approximate GPs to consider non-spatial effects.

Exceptionally, the integrated nested Laplace approximation (INLA) developed by Rue et al. (2009) and the fast AMM framework developed by Wood (2008; 2011), Wood et al. (2015; 2017), Li and Wood (2019), allow us to estimate spatial and non-spatial effects from large samples. INLA provides computationally efficient estimation and inference for latent Gaussian models, which are Bayesian AMMs. Although INLA is indeed flexible and useful for large samples, it still has the following limitations: (i) its computational complexity grows exponentially with respect to the number of hyperparameters (Rue et al., 2009); and, (ii) its memory consumption depends on $N$. Thus, INLA is not suitable for cases with very large samples (e.g., $N = 1,000,000$) and many hyperparameters (e.g., 10).

By contrast, the fast AMM approach, which does not suffer from problems (i) and (ii), is extremely fast and memory efficient (see Wood 2015; Wood and Shaddick 2017). Moreover, the geo-additive (mixed) model (GeoAMM; Kammann and Wand 2003), which is an AMM that considers spatial dependence using a rank reduced GP, is available under the framework. Because this AMM framework allows for constricting the (restricted) log-likelihood through block-wise data processing, the memory consumption is independent of $N$. Furthermore, because this approach eliminates matrices whose size depends on $N$ from the log-likelihood function a priori, the maximum likelihood (ML), or the restricted ML (REML) maximization is fast even for very large samples. The fast AMM, especially GeoAMM, is useful for large-scale spatial regression modeling.

A limitation of GeoAMM is that it cannot estimate the spatial structure behind residuals and other model elements such as SVCs. Because a misspecification of spatial structure can result in erroneous conclusions in regression analysis, as repeatedly suggested in spatial econometrics (e.g., Anselin 2003; LeSage and Pace 2009), AMM must be extended to estimate spatial structure while minimizing additional computational loads. Another problem is the degeneracy of the rank reduced GP in GeoAMM. Although 20 – 40 basis functions are often considered as sufficient in studies on GeoAMM (e.g., Kneib et al. 2009), more basis functions might be needed to mitigate degeneracy of the rank reduced GP (see Stein 2014). For example, Murakami and Griffith (2019a) suggest that 200 spatial basis functions are required to eliminate residual spatial dependence when $N$ = 80,000. To consider a large number of basis functions in a large-scale spatial regression analysis, memory efficiency is crucially important.

Given this background, this study develops a memory-free AMM estimating spatially dependent effects with unknown scales and other non-spatial effects. This development is done by extending Murakami and Griffith (2019b). Their focus is limited to an SVC model, which is a particular AMM. In addition, the memory consumption of their approach, which stores spatial basis functions, grows with respect to $N$. Therefore, their approach cannot handle very large $N$ that exceeds the memory limit. In other words, their approach is inferior to the fast AMM in terms of both generality and computational efficiency. To overcome these two limitations, we generalize their approach to a spatial AMM approach with no memory limitations.

The subsequent sections are organized as follows. Section 2 introduces our model and estimation method. Section 3 introduces a fast estimation procedure. Section 4 compares it with alternatives through Monte Carlo simulation experiments. Finally, Section 5 concludes our discussion.

## 2. Spatial additive mixed model

Section 2.1 introduces the linear AMM, which we consider, whereas Sections 2.2 to 2.4 introduce its sub-models describing spatial and non-spatial effects. Specifically, after introducing a typical model to describe spatial effects in Section 2.2, Section 2.3 introduces our spatial modeling approach based on the Moran coefficient (MC; Moran, 1950). Subsequently, Section 2.4 explains models that describe non-spatial effects.

2.1. Model

This study considers the following linear AMM:

$$\mathbf{y} = \mathbf{X}\mathbf{b} + \sum_{p=1}^{P} \mathbf{A}_p \mathbf{V}(\boldsymbol{\theta}_p)\mathbf{u}_p + \boldsymbol{\varepsilon} \qquad \mathbf{u}_p \sim N(\mathbf{0}_{L_p}, \sigma^2 \mathbf{I}_{L_p}) \qquad \boldsymbol{\varepsilon} \sim N(\mathbf{0}, \sigma^2 \mathbf{I}) \qquad (1)$$

where **y** is a vector of response values ($N \times 1$) observed in a study region $D \subset R^2$, $\mathbf{X} = [\mathbf{x}_1, \cdots, \mathbf{x}_K]$ is a matrix of explanatory variables ($N \times K$), and **b** is a $K \times 1$ vector of regression coefficients. $\mathbf{0}_{L_p}$ ($L_p \times 1$) e **0** ($N \times 1$) are vectors of zeros, $\mathbf{I}_{L_p}$ ($L_p \times L_p$) and **I** ($N \times N$) are identity matrices, and $\sigma^2$ is a variance parameter. The additive term $f(\mathbf{z}_p) = \mathbf{A}_p \mathbf{V}(\boldsymbol{\theta}_p) \mathbf{u}_p$ models the effects from $p$-th covariate $\mathbf{z}_p$ using a matrix $\mathbf{A}_p$ ($N \times L_p$) of basis functions generated based on $\mathbf{z}_p$, a matrix $\mathbf{V}(\boldsymbol{\theta}_p)$ ($L_p \times L_p$) estimating the influence from each basis function, and a set of variance parameters $\boldsymbol{\theta}_p$. While we assumed the same variances for $\mathbf{u}_p$s and $\boldsymbol{\varepsilon}$ for ease of explanation later, the variance of $\mathbf{u}_p$ is readily rescaled by assuming a variance parameter inside $\mathbf{V}(\boldsymbol{\theta}_p)$. The standard AMM equals Eq.(1) with $\mathbf{V}(\boldsymbol{\theta}_p) = \tau_p^2 \mathbf{V}_p$, where $\mathbf{V}_p$ is a known matrix and $\tau_p^2$ is a parameter. This study assumes parameters $\boldsymbol{\theta}_p$ inside the $\mathbf{V}_p$ matrix to estimate the scale of spatial process.

Below, Section 2.2 explains a typical specification to spatial dependence modeling in AMM studies. Then, Sections 2.2 and 2.3 explain our specifications for $\mathbf{A}_p \mathbf{V}(\boldsymbol{\theta}_p) \mathbf{u}_p$ to model spatial and non-spatial effects, respectively.

2.2. Modeling spatial dependence: A typical approach

This study models spatial dependence by specifying $\mathbf{A}_1 \mathbf{V}(\boldsymbol{\theta}_1) \mathbf{u}_1$ as a spatial process. Although the full rank GP, which is widely used for spatial dependence modeling, is computationally expensive (see Section 1), GeoAMM typically approximates a GP using $L_1$ ($<N$) basis functions and inducing points (or knot/anchor points), which are distributed in the study region by a space-filling algorithm, characterizing these bases (Kammann and Wand 2003). The rank $L_1$ GP, which is derived by minimizing the predicted error variance, yields

$$f(\mathbf{z}_1) = \tau_1^2 \mathbf{C}_{L_1} \mathbf{C}_{L_1 L_1}^{-1} \mathbf{u}_1, \quad \mathbf{u}_1 \sim N(\mathbf{0}_{L_1}, \sigma^2 \mathbf{I}_{L_1}) \tag{2}$$

which implies $\mathbf{A}_1 = \mathbf{C}_{L_1}$ and $\mathbf{V}(\boldsymbol{\theta}_1) = \tau_1^2 \mathbf{C}_{L_1 L_1}^{-1}$ (where $\mathbf{z}_1$ represents the spatial coordinates). $\mathbf{C}_{L_1}$ is a $N \times L_1$ matrix whose $(i, l_1)$-th element equals $c(d_{i,l_1})$, which is a known distance decay function with respect to the Euclidean metric $d_{i,l_1}$ between $i$-th sample site and $l_1$-th inducing point, and $\mathbf{C}_{L_1 L_1}$ is a $L_1 \times L_1$ matrix whose $(l_1, l'_1)$-th element equals $c(d_{l_1, l'_1})$. Later, this study assumes an exponential kernel $c(d_{i,l_1}) = \exp(-d_{i,l_1}/r)$, where $r$ is a scaling parameter. The spatial process has a large-scale/long-range spatial pattern if $r$ is large, whereas the opposite is true for small $r$.

A limitation of this specification is that $c(d_{i,l_1})$ is assumed to be known. In other words, $r$ must be given a priori despite that the estimation of $r$ is needed to avoid misspecification of spatial scale that can lead to erroneous results (Murakami et al. 2019). Unfortunately, an unknown $r$, which implies an unknown $\mathbf{A}_1 = \mathbf{C}_{L_1}$, is somewhat inconsistent with the usual assumption of AMM. Furthermore, it is difficult to apply the fast AMM of Wood et al. (2015) if $r$ is unknown. He applies the QR decomposition for $\mathbf{A}_1 = \mathbf{C}_{L_1}$. If $\mathbf{A}_1$ is known, the decomposition is required only once before parameter estimation. However, if $\mathbf{A}_1$ changes depending on the unknown $r$, the QR-decomposition must be iterated during the estimation with this computation being slow. The approach of Wood et al.

(2015) is not suitable in our case.

2.3. Modeling spatial dependence: Our approach

This section introduces a MC-based approach as an alternative of the typical approach introduce in Section 2.2. Section 2.3.1 explains the MC. Sections 2.3.2 defines a MC-based spatial process, and Section 2.3.3 approximates this process for fast approximation. Then, Section 2.3.4 formulates this approximate spatial process as an element of the AMM.

2.3.1. Moran coefficient

Instead of the typical GeoAMM specification, we rely on a specification based on the MC, which is a spatial dependence diagnostic statistic. As explained later, the MC is interpretable as a scale parameter. Using this property, we attempt to replace the estimation of the range parameter $r$ with an estimation of the MC value.

The MC for a random variable vector $\mathbf{r}$ is defined as

$$MC[\mathbf{r}] = \frac{N}{\mathbf{1}'\mathbf{C}_0\mathbf{1}} \frac{\mathbf{r}'\mathbf{M}\mathbf{C}_0\mathbf{M}\mathbf{r}}{\mathbf{r}'\mathbf{M}\mathbf{r}}, \qquad (3)$$

where $"\,'\,"$ represents the matrix transpose, $\mathbf{C}_0$ is a known spatial proximity matrix ($N \times N$) with zero diagonals and $\mathbf{M} = \mathbf{I} - \mathbf{1}\mathbf{1}'/N$ is a centering matrix. $MC[\mathbf{r}]$ indicates a positive value if the elements of $\mathbf{r}$ are positively spatially dependent (i.e., neighborhood values are similar), a negative value if they are negatively dependent (i.e., neighborhood values are dissimilar), and approaches $-1/(N-1) \approx 0$ if they are independently distributed (see Griffith 2003).

2.3.2. A Moran coefficient-based spatial process

This study uses the Moran eigenvectors (or Moran basis: Griffith 2003; Hughes and Haran 2013; Murakami and Griffith 2015) that are spatial basis functions interpretable based on the MC. Let us eigen-decompose the double centered proximity matrix $\mathbf{MC_0M}$ as $\mathbf{E\Lambda E'}$. Moran bases are defined by the eigenvectors $\mathbf{E} = [\mathbf{e}_1, \cdots, \mathbf{e}_N]$ that are arranged in descending order of their corresponding eigenvalues $\{\lambda_1, \cdots, \lambda_N\}$ that comprise a $L \times L$ diagonal matrix $\mathbf{\Lambda}$. The MC value for the $l$-th eigenvalue yields

$$MC[\mathbf{e}_l] = \frac{N}{\mathbf{1'C_0 1}} \frac{\mathbf{e}_l' \mathbf{MC_0 M e}_l}{\mathbf{e}_l' \mathbf{M e}_l} = \frac{N}{\mathbf{1'C_0 1}} \frac{\mathbf{e}_l' \mathbf{E\Lambda E' e}_l}{\mathbf{e}_l' \mathbf{e}_l} = \frac{N}{\mathbf{1'C_0 1}} \lambda_l. \qquad (4)$$

Eq. (4) suggests that the Moran eigenvectors corresponding to positive eigenvalues explain positive spatial dependence, whereas the opposite is true for negative spatial dependence. Thus, the Moran eigenvectors are basis functions describing latent spatial dependence, with each level being indexed by the MC value (Griffith 2003). It is known that $\mathbf{e}_1$ has the largest-scale map pattern, and $\mathbf{e}_l$ has the $l$-th largest-scale map pattern that is orthogonal to $\{\mathbf{e}_1, \cdots, \mathbf{e}_{l-1}\}$. This implies that the MC takes a large positive value in the presence of large-scale spatial dependence, and a small value in the presence of small-scale spatial variations. In other words, MC is an indicator of the scale of spatial dependence.

Moran eigenvectors are interpretable based on a GP. To see this, consider a GP $\widetilde{\mathbf{w}} \sim N(\mathbf{0}, \mathbf{C})$, where $\mathbf{C} = \mathbf{C}_0 + \mathbf{I}$. The GP can be expanded as $\widetilde{\mathbf{w}} = \mu_{\widetilde{\mathbf{w}}} \mathbf{1} + \widetilde{\mathbf{w}}_c$, where $\mu_{\widetilde{\mathbf{w}}} = \mathbf{1}'\widetilde{\mathbf{w}}/N$ and $\widetilde{\mathbf{w}}_c \sim N(\mathbf{0}, \mathbf{MC}_0\mathbf{M} + \mathbf{MM})$, which is a centered GP. The $\mathbf{MM}(= \mathbf{MIM})$ matrix is a doubly centered identity matrix explaining independent variations. Thus, $\mathbf{MC}_0\mathbf{M}$ or $\mathbf{E\Lambda E}'$ explains pure spatial dependence effects extracted by centering and excluding the independent variations from a GP. Thus, the Moran eigenvectors are interpretable in terms of both the MC and GP.

We use the Moran eigenvectors because of the interpretability and the effectiveness of spatial dependence modeling, which has been reported by Griffith and Peres-Neto (2006), and Tiefelsdorf and Griffith (2007), among others. Following Dray et al. (2006) and Hughes and Haran (2013), we use the $L$ eigenvectors corresponding to positive eigenvalues $\{\lambda_1, \cdots, \lambda_L\}$ because positive spatial dependence is dominant in regional and natural science (Griffith 2003). The resulting rank $L$ spatial process is formulated as

$$\mathbf{w} = \left(\frac{\tau^2}{\sigma^2}\right) \mathbf{E\Lambda}^{\alpha} \mathbf{u} \qquad \mathbf{u} \sim N(\mathbf{0}_L, \sigma^2 \mathbf{I}_L), \tag{5}$$

which equals an approximate GP $\mathbf{w} \sim N(\mathbf{0}, \tau^2 \mathbf{E\Lambda}^{2\alpha} \mathbf{E})$. $\mathbf{I}_L$ is an $L \times L$ identity matrix and $\mathbf{0}_L$ is a $L \times 1$ vector of zeros, and $\tau^2 \in [0, \infty)$ and $\alpha \in (-\infty, \infty)$ are parameters.[1]

Although the interpretation of $\alpha$ has never been studied, it acts as a parameter determining

---

[1] $\mathbf{w}$ yields a negatively dependent spatial process if $\mathbf{\Lambda}$ is defined using the absolute values of negative eigenvalues and $\mathbf{E}$ is defined using their corresponding eigenvectors. Our approach is capable of analyzing negative spatial dependence situations (see Griffith, 2006).

the MC value of the process. To see this, let us evaluate the expectation of $MC[\mathbf{w}]$ as follows:

$$E\left[MC\left[\left(\frac{\tau^2}{\sigma^2}\right)\mathbf{E}\Lambda^\alpha\mathbf{u}\right]\right] = \frac{N}{\mathbf{1}'\mathbf{C}_0\mathbf{1}}E\left[\frac{\mathbf{u}'\Lambda^\alpha\mathbf{E}'\mathbf{MC}_0\mathbf{ME}\Lambda^\alpha\mathbf{u}}{\mathbf{u}'\Lambda^\alpha\mathbf{E}'\mathbf{ME}\Lambda^\alpha\mathbf{u}}\right] = \frac{N}{\mathbf{1}'\mathbf{C}_0\mathbf{1}}\frac{E[\mathbf{u}'\Lambda^{2\alpha+1}\mathbf{u}]}{E[\mathbf{u}'\Lambda^{2\alpha}\mathbf{u}]}$$
$$= \frac{N}{\mathbf{1}'\mathbf{C}_0\mathbf{1}}\frac{\sum_{l=1}^{L}\lambda_l^{2\alpha+1}}{\sum_{l=1}^{L}\lambda_l^{2\alpha}}.\tag{6}$$

Eq.(6) suggests that the MC value of the spatial process is determined solely by the $\alpha$ parameter. It can be further expanded as

$$E[MC[\mathbf{w}]] = \frac{N}{\mathbf{1}'\mathbf{C}_0\mathbf{1}}\frac{\lambda_1 + \sum_{l=2}^{L}\frac{\lambda_l^{2\alpha}}{\lambda_1^{2\alpha}}\lambda_l}{1 + \sum_{l=2}^{L}\frac{\lambda_l^{2\alpha}}{\lambda_1^{2\alpha}}}.\tag{7}$$

Eq.(7) implies that, as $\alpha$ increases, $E[MC[\mathbf{w}]]$ asymptotically converges to

$$\lim_{\alpha \to \infty} E[MC[\mathbf{w}]] = \frac{N}{\mathbf{1}'\mathbf{C}_0\mathbf{1}}\lambda_1,\tag{8}$$

which equals the maximum possible MC value (see Griffith 2003). Likewise, as $\alpha$ decreases, $E[MC[\mathbf{w}]]$ converges to

$$\lim_{\alpha \to -\infty} E[MC[\mathbf{w}]] = \frac{N}{\mathbf{1}'\mathbf{C}_0\mathbf{1}}\lambda_L,\tag{9}$$

which equals the smallest possible MC value in our setting, which is a value near zero (recall that $\lambda_L$ is the smallest positive eigenvalue). Moreover, $\alpha$ is a parameter estimating the MC value of the process, whereas $\tau^2$ estimates the variance of the process. Because the MC value indicates the scale of spatial dependence, the $\alpha$ parameter acts as a scale parameter. Although it is usual to estimate spatial scale using a range parameter $r$ inside $\mathbf{C}_0$, the relationship between the range parameter value and the MC value is unclear. Thus, we prefer the $\alpha$-based specification, which has a clear interpretation, and, as we explain subsequently, allows us to estimate scales computationally efficiently.

Note that our specification requires the base kernel matrix $\mathbf{C}_0$ to evaluate the MC value.

Later, we define the $(i, j)$-th element of the $\mathbf{C}_0$ matrix by $c(d_{i,j}) = \exp(-d_{i,l_1}/r)$, where $r$ is the maximum distance in the minimum spanning tree connecting sample sites because this criterion is widely accepted, especially in ecological studies (e.g., Dray et al. 2006). Murakami and Griffith (2019a) showed that the accuracy of spatial process modeling holds even if the exponential kernel is replaced with Gaussian or spherical kernels. Thus, the exponential kernel may be replaced with the Gaussian, spherical, or, potentially, other kernels. They also showed that the Moran eigenvector approach is robust against the misspecification of $r$, and that this robustness holds even if the eigenvectors are approximated in the way introduced in the next section. Although Murakami and Griffith (2019a) also verified the accuracy for large samples using a symmetric $\mathbf{C}_0$ matrix,[2] Kelejian and Prucha (1998) and Arbia et al. (2019) suggested inaccuracies occur with the eigen-decomposition for asymmetric spatial proximity matrices. We assume $\mathbf{C}$ as symmetric.

2.3.3. Approximation for the MC-based process

Although the eigen-decomposition, whose complexity equals $O(N^3)$, is intractable for large samples, its complexity can be reduced to a linear order with respect to $N$ by applying an approach based on the Nystrom approximation (Dreneas and Mahoney 2005). The approximated Moran eigenvectors and eigenvalues yield (see Murakami and Griffith 2019a):

$$\widehat{\mathbf{E}} = \left[\mathbf{C}_{NL} - \mathbf{1} \otimes \left(\frac{\mathbf{1}'_L \mathbf{C}_L}{L}\right)\right] \mathbf{E}_L (\mathbf{\Lambda}_L + \mathbf{I}_L)^{-1},$$
$$\widehat{\mathbf{\Lambda}} = \frac{N+L}{L} (\mathbf{\Lambda}_L + \mathbf{I}_L) - \mathbf{I}_L, \tag{10}$$

where $\otimes$ is the Kroncker product operator, and $\mathbf{C}_{NL}$ is a $N \times L$ matrix whose $(i,l)$-th element equals a given kernel $c(d_{i,l}) = \exp(-d_{i,l}/r)$, $\mathbf{E}_L \mathbf{\Lambda}_L \mathbf{E}'_L = \mathbf{M}_L \mathbf{C}_L \mathbf{M}_L$, where $\mathbf{C}_L$ is a $L \times L$ proximity matrix among $L$ inducing points distributed across the study area, and $\mathbf{M}_L = \mathbf{I}_L - \mathbf{1}_L \mathbf{1}'_L/L$. Following Zhang and Kwok (2010), the inducing points are defined by $k$-means clustering centers.

---

[2] They confirmed that the Moran eigenvector approach successfully captures/eliminates residual spatial dependence for large samples up to $N = 80,000$.

Unlike the usual eigen-decomposition, this approximation allows for evaluating the eigenvectors sequentially for $H$ sub-samples, which are indexed by $h \in \{1, ..., H\}$. The subset of the $L$ eigenvectors for the $h$-th sub-samples is derived by replacing $\mathbf{C}_{NL}$ with a proximity matrix between the $L$ inducing points and the samples in the $h$-th subset (and $\mathbf{1}$ with a vector of ones with appropriate length). This property is used for memory saving. Namely, we assume using the approximate eigen-pairs $\widehat{\mathbf{E}}$ and $\widehat{\mathbf{\Lambda}}$.

2.3.4. The MC-based process as an element of the AMM

The MC-based process $\mathbf{w} = \left(\frac{\tau^2}{\sigma^2}\right) \widehat{\mathbf{E}} \widehat{\mathbf{\Lambda}}^{\alpha} \mathbf{u}$ is easily incorporated into the AMM Eq.(1) by specifying $\mathbf{A}_p \mathbf{V}(\boldsymbol{\theta}_p) \mathbf{u}_p$ using $\mathbf{A}_p = \widehat{\mathbf{E}}$, $\mathbf{V}(\boldsymbol{\theta}_p) = \left(\frac{\tau^2}{\sigma^2}\right) \widehat{\mathbf{\Lambda}}^{\alpha}$, and $\mathbf{u}_p = \mathbf{u}$. In addition, the MC-based process is available to estimate SVCs. In this case, $\mathbf{A}_p\mathbf{V}(\boldsymbol{\theta}_p)\mathbf{u}_p$ is given using $\mathbf{A}_p = \mathbf{x}_p°\hat{\mathbf{E}}$ and $\mathbf{V}(\boldsymbol{\theta}_p) = \left(\frac{\tau_p^2}{\sigma^2}\right)\hat{\boldsymbol{\Lambda}}^{\alpha_p}$, and $\mathbf{u}_p = \mathbf{u}$, where ° in this study is an operator multiplying the vector in the left side with each column of the matrix in the right side of the operator. Then, given $k = p$, our AMM Eq(1) yields the following SVC model (see Murakami et al. 2017):

$$\mathbf{y} = \sum_{p=1}^{P}\mathbf{x}_p°(b_p\mathbf{1} + \mathbf{w}_p) + \boldsymbol{\varepsilon} \qquad \boldsymbol{\varepsilon} \sim N(\mathbf{0}, \sigma^2\mathbf{I})$$

$$\mathbf{w}_p = \left(\frac{\tau_p^2}{\sigma^2}\right)\hat{\mathbf{E}}\hat{\boldsymbol{\Lambda}}^{\alpha_p}\mathbf{u} \qquad \mathbf{u}_p \sim N(\mathbf{0}, \sigma^2\mathbf{I}_p)$$

(11)

where $b_p$ is the $p$-th element of $\mathbf{b}$. $b_p\mathbf{1} + \mathbf{w}_p$ is the SVC where the first and second terms denote the constant mean component and the spatially varying component, respectively. In short, our AMM is useful for spatial modeling just like GeoAMM. However, unlike GeoAMM, our specification estimates the scale of each spatial process using the $\{\alpha, \alpha_1, \dots, \alpha_P\}$ parameters measuring the MC values, which is interpretable as an indicator of scale underlying each process.

2.4. Modeling non-spatial effects

A wide variety of terms characterizing non-spatial effects are modeled by specifying the term $\mathbf{A}_p\mathbf{V}(\boldsymbol{\theta}_p)\mathbf{u}_p$. They include (a) group effects ($\mathbf{A}_p$: a matrix of dummy variables indicating groups, and $\boldsymbol{\Sigma}(\boldsymbol{\theta}_p) = \tau_p^2\mathbf{I}_p$), (b) smooth effects ($\mathbf{A}_p$: a matrix consists of smoothing splines, and $\boldsymbol{\Sigma}(\boldsymbol{\theta}_p) = \tau_p^2\mathbf{I}_p$), and (c) temporal effects (see Hodges 2016).

Most studies in geostatistics and spatial econometrics ignore (a) and (b). However, if these non-spatial variations are ignored, the ignored variations might blue spatial effects, and make their identification difficult. Such a difficulty might be severe in our case, which considers spatial effects underlying not only residuals but also regression coefficients. A consideration of non-spatial effects is needed to appropriately estimate them.

Consideration of group effects is especially important because its ignorance can lead to erroneous inference, which is the so called Simpson's paradox (e.g., Samuels 1993). This paradox states that if individual-level trends and group-level trends have differences, ignoring the latter leads to biased inference for the individuals. For instance, when analyzing the impact of an educational program on school achievement of individual students, we need to consider not only individual-level trends but also school-level trends. Following Hox (1998) and Goldstein (2011), among others, who suggest the importance of incorporating group effects, we consider these effects later.

In summary, Section 2 introduces MC-based AMM emphasizing spatial modeling. The next section explains how to estimate the AMM computationally efficiently.

## 3. Estimation

This section develops a fast and memory-free REML for the MC-based AMM. Section 3.1 introduces our restricted log-likelihood function. Section 3.2 introduces its fast maximization procedure. Furthermore, in Section 3.3, this procedure is extended to a memory-free procedure for very large samples.

## 3.1. The restricted log-likelihood

This study estimates the spatial AMM by maximizing the Type II restricted log-likelihood. Unlike the standard Type I likelihood that treats latent variables as pseudo-observations, the Type II likelihood integrates out latent variables by regarding them as nuisance. The Type II log-likelihood for our AMM Eq.(1) is defined as $\log p(\mathbf{y}|\mathbf{b},\Theta) = \log \int p(\mathbf{y},\tilde{\mathbf{u}}|\mathbf{b},\Theta)p(\tilde{\mathbf{u}})d\tilde{\mathbf{u}}$, where $\tilde{\mathbf{u}} \in \{\mathbf{u}_1, \cdots \mathbf{u}_P\}$ and $\Theta \in \{\boldsymbol{\theta}_1, \cdots \boldsymbol{\theta}_P, \sigma^2\}$, whereas the restricted log-likelihood is formulated as $loglik_R(\Theta) = \log \int p(\mathbf{y}|\mathbf{b},\Theta)d\mathbf{b}$. Bates (2010) derives the closed-form expression of the Type II restricted log-likelihood for the standard linear mixed model (LMM). Because our model Eq.(1) is identical to the LMM, our Type II restricted log-likelihood is formulated as follows, just like the Bates's likelihood:

$$loglik_R(\Theta) = -\frac{1}{2}\ln|\mathbf{R}(\Theta)| - \frac{N-K}{2}\left(1 + \ln\left(\frac{2\pi d(\Theta)}{N-K}\right)\right), \quad (12)$$

where

$$d(\Theta) = \left\|\mathbf{y} - \mathbf{X}\hat{\mathbf{b}} - \sum_{p=1}^{P}\mathbf{A}_p\mathbf{V}(\boldsymbol{\theta}_p)\hat{\mathbf{u}}_p\right\|^2 + \sum_{p=1}^{P}\|\hat{\mathbf{u}}_p\|^2, \quad (13)$$

$$\begin{bmatrix}\hat{\mathbf{b}}\\\hat{\mathbf{u}}_1\\\vdots\\\hat{\mathbf{u}}_P\end{bmatrix} = \mathbf{R}(\Theta)^{-1}\begin{bmatrix}\mathbf{X}'\mathbf{y}\\\mathbf{V}(\boldsymbol{\theta}_1)\mathbf{A}'_1\mathbf{y}\\\vdots\\\mathbf{V}(\boldsymbol{\theta}_P)\mathbf{A}'_P\mathbf{y}\end{bmatrix}. \quad (14)$$

$$\mathbf{R}(\Theta) = \begin{bmatrix}\mathbf{X}'\mathbf{X} & \mathbf{X}'\mathbf{A}_1\mathbf{V}(\boldsymbol{\theta}_1) & \cdots & \mathbf{X}'\mathbf{A}_P\mathbf{V}(\boldsymbol{\theta}_P)\\\mathbf{V}(\boldsymbol{\theta}_1)\mathbf{A}'_1\mathbf{X} & \mathbf{V}(\boldsymbol{\theta}_1)\mathbf{A}'_1\mathbf{A}_1\mathbf{V}(\boldsymbol{\theta}_1)+\mathbf{I} & \cdots & \mathbf{V}(\boldsymbol{\theta}_1)\mathbf{A}'_1\mathbf{A}_P\mathbf{V}(\boldsymbol{\theta}_P)\\\vdots & \vdots & \ddots & \vdots\\\mathbf{V}(\boldsymbol{\theta}_P)\mathbf{A}'_P\mathbf{X} & \mathbf{V}(\boldsymbol{\theta}_P)\mathbf{A}'_P\mathbf{A}_1\mathbf{V}(\boldsymbol{\theta}_1) & \cdots & \mathbf{V}(\boldsymbol{\theta}_P)\mathbf{A}'_P\mathbf{A}_P\mathbf{V}(\boldsymbol{\theta}_P)+\mathbf{I}\end{bmatrix}. \quad (15)$$

In our REML, the variance parameters $\{\boldsymbol{\theta}_1, \cdots \boldsymbol{\theta}_P\}$ can be estimated by numerically maximizing the likelihood Eq. (12). Then, the coefficients $[\hat{\mathbf{b}}', \hat{\mathbf{u}}'_1, \cdots, \hat{\mathbf{u}}'_P]'$ characterizing spatial effects are estimated by substituting the estimated variance parameters into Eq. (12).

Standard errors of coefficients are useful for significance testing. Although a standard error quantifies estimation error, the random effects estimates $\{\hat{\mathbf{u}}_1, \cdots, \hat{\mathbf{u}}_P\}$ include variance from not only the estimation error, but also from the randomness of the effects $\{\mathbf{u}_1, \cdots, \mathbf{u}_P\}$ themselves (see Hodges, 2014). To evaluate pure estimation error, the standard error of $\{\hat{\mathbf{b}} - \mathbf{b}, \hat{\mathbf{u}}_1 - \mathbf{u}_1, \cdots, \hat{\mathbf{u}}_P - \mathbf{u}_P\}$, which equals $\{\hat{\mathbf{b}}, \hat{\mathbf{u}}_1 - \mathbf{u}_1, \cdots, \hat{\mathbf{u}}_P - \mathbf{u}_P\}$, must be evaluated, where $\mathbf{b}$ is a vector of true fixed coefficients. Based on Henderson, (1975; 1984), the variance-covariance matrix is $Var\begin{bmatrix}\hat{\mathbf{b}} - \mathbf{b} \\ \hat{\mathbf{u}}_1 - \mathbf{u}_1 \\ \vdots \\ \hat{\mathbf{u}}_P - \mathbf{u}_P\end{bmatrix} = \sigma^2 \mathbf{R}(\mathbf{\Theta})^{-1}$.

Consider a subset $\begin{bmatrix}\hat{b}_k - b_k \\ \hat{\mathbf{u}}_p - \mathbf{u}_p\end{bmatrix}$ of the coefficients. The variance is specified as $Var\begin{bmatrix}\hat{b}_k - b_k \\ \hat{\mathbf{u}}_p - \mathbf{u}_p\end{bmatrix} = \sigma^2 \mathbf{R}(\mathbf{\Theta})^*_{k.p}$ where $\mathbf{R}(\mathbf{\Theta})^*_{k.p}$ is a subset of $\mathbf{R}(\mathbf{\Theta})^{-1}$. The standard error of spatial or non-spatial effects, which are defined by a linear combination of fixed and/or random effects, and thus can be written as $\hat{b}_k \mathbf{1} + \mathbf{A}_p \mathbf{V}(\mathbf{\theta}_p) \hat{\mathbf{u}}_P$, is evaluated by taking the square root of $Var[(\hat{b}_k \mathbf{1} + \mathbf{A}_p \mathbf{V}(\mathbf{\theta}_p) \hat{\mathbf{u}}_P) - (b_k \mathbf{1} + \mathbf{A}_p \mathbf{V}(\mathbf{\theta}_p) \mathbf{u}_P)] = \hat{\sigma}^2 [\mathbf{1}, \mathbf{A}_p \mathbf{V}(\mathbf{\theta}_p)] \mathbf{R}(\mathbf{\Theta})^*_{k.p} \begin{bmatrix}\mathbf{1}' \\ \mathbf{V}(\mathbf{\theta}_p) \mathbf{A}'_p\end{bmatrix}$.[3]

Unlike the standard spatial (or GP) models, our likelihood does not include the inverse of a $N \times N$ matrix that implies computational complexity of $O(N^3)$ and memory consumption of $O(N^2)$.

---

[3] Our MC-based spatial process yields a rank reduced GP (see Murakami et al., 2017). Therefore, the standard error of the MC-based process can be underestimated if the full GP is regarded as the true process. Murakami and Griffith (2019) showed that standard error bias is small if the true process is a positively dependent spatial one, in which the MC-based approach intended. For fitting a full GP, bias correction (see e.g., Finley et al., 2009) will be an important future step.

Thus, ours is considerably faster than the standard specification. Still, our REML is less efficient in terms of both computational efficiency and memory usage. Regarding computational efficacy, $\mathbf{R}(\mathbf{\Theta})^{-1}$ and $|\mathbf{R}(\mathbf{\Theta})|$, whose complexities are both $O((K + K\sum_{p=1}^{P} L_p)^3)$, must be repeatedly evaluated to find the REML estimates for $\mathbf{\Theta}$; $\mathbf{R}(\mathbf{\Theta})$ must be constructed in each iteration. The REML is slow if N and/or P is large. Regarding memory consumption, our likelihood function includes large matrices $\{\mathbf{A}_1, \cdots, \mathbf{A}_P\}$ that cannot be stored if N is large, say, in the millions.

To address these problems, Sections 3.2 and 3.3 extend REML to the case of big data. For fast variance parameter estimation, we employ a procedure developed by Murakami and Griffith (2019b), which is explained in Section 3.2. For memory saving, we newly develop a memory efficient procedure in Section 3.3.

3.2. Fast parameter estimation

Eq. (13) - (15) have the following expression:

$$d(\mathbf{\Theta}) = m_{y,y} - 2[\hat{\mathbf{b}}', \hat{\mathbf{u}}'_1, \cdots \hat{\mathbf{u}}'_P] \begin{bmatrix} \mathbf{m}_0 \\ \mathbf{V}(\mathbf{\theta}_1)\mathbf{m}_1 \\ \vdots \\ \mathbf{V}(\mathbf{\theta}_P)\mathbf{m}_P \end{bmatrix} + [\hat{\mathbf{b}}', \hat{\mathbf{u}}'_1, \cdots \hat{\mathbf{u}}'_P] \mathbf{P}_0 \begin{bmatrix} \hat{\mathbf{b}} \\ \hat{\mathbf{u}}_1 \\ \vdots \\ \hat{\mathbf{u}}_P \end{bmatrix} + \sum_{p=1}^{P} \|\hat{\mathbf{u}}_P\|^2, \quad (16)$$

$$\begin{bmatrix} \hat{\mathbf{b}} \\ \hat{\mathbf{u}}_1 \\ \vdots \\ \hat{\mathbf{u}}_P \end{bmatrix} = \mathbf{R}(\mathbf{\Theta})^{-1} \begin{bmatrix} \mathbf{m}_0 \\ \mathbf{V}(\mathbf{\theta}_1)\mathbf{m}_1 \\ \vdots \\ \mathbf{V}(\mathbf{\theta}_P)\mathbf{m}_P \end{bmatrix}, \quad (17)$$

$$\mathbf{R}(\mathbf{\Theta}) = \begin{bmatrix} \mathbf{M}_{0,0} & \mathbf{M}_{0,1}\mathbf{V}(\mathbf{\theta}_1) & \cdots & \mathbf{M}_{0,P}\mathbf{V}(\mathbf{\theta}_P) \\ \mathbf{V}(\mathbf{\theta}_1)\mathbf{M}_{1,0} & \mathbf{V}(\mathbf{\theta}_1)\mathbf{M}_{1,1}\mathbf{V}(\mathbf{\theta}_1) + \mathbf{I} & \cdots & \mathbf{V}(\mathbf{\theta}_1)\mathbf{M}_{1,P}\mathbf{V}(\mathbf{\theta}_P) \\ \vdots & \vdots & \ddots & \vdots \\ \mathbf{V}(\mathbf{\theta}_P)\mathbf{M}_{P,0} & \mathbf{V}(\mathbf{\theta}_P)\mathbf{M}_{P,1}\mathbf{V}(\mathbf{\theta}_1) & \cdots & \mathbf{V}(\mathbf{\theta}_P)\mathbf{M}_{P,P}\mathbf{V}(\mathbf{\theta}_P) + \mathbf{I} \end{bmatrix}, \quad (18)$$

where $m_{y,y} = \mathbf{y}'\mathbf{y}$, $\mathbf{m}_0 = \mathbf{X}'\mathbf{y}$, $\mathbf{m}_p = \mathbf{A}'_p\mathbf{y}$, $\mathbf{M}_{0,0} = \mathbf{X}'\mathbf{X}$, $\mathbf{M}_{0,p} = \mathbf{X}'\mathbf{A}_p$, and $\mathbf{M}_{p.p'} = \mathbf{A}'_p\mathbf{A}_{p'}$. Eq. (16) – (18) suggest that, if these inner product matrices are evaluated a priori, the restricted log-likelihood does not include any matrices whose size depends on the sample size. Thus, the computational cost to estimate $\Theta$ is independent of $N$.

Still, the iterative evaluations of $\mathbf{R}(\Theta)^{-1}$ and $|\mathbf{R}(\Theta)|$, with complexities being $O((K + K\sum_{p=1}^{P} L_p)^3)$ respectively, are needed to maximize $loglik_R(\Theta)$ with respect to $\Theta$. Fortunately, Murakami and Griffith (2019b) derive an approach to lighten the computational cost. In their approach, the variance parameters $\{\theta_1, \cdots \theta_P\}$ are sequentially updated until convergence. When estimating $\theta_P$, $loglik_R(\theta_p)|\Theta_{-p}$, is maximized by partially updating $\mathbf{R}(\Theta)^{-1}$ and $|\mathbf{R}(\Theta)|$ where $\Theta_{-p} \in \{\theta_1, \cdots, \theta_{p-1}, \theta_{p+1}, \cdots, \theta_P\}$. If matrices that are independent of $\theta_P$ are processed a priori, the resulting computational cost to evaluate $loglik_R(\theta_p)|\Theta_{-p}$ is only $O(L_p^3)$, which is independent of $N$ and $P$. Eventually, our REML estimation is fast even if $N$ and $P$ are very large (see Murakami and Griffith, 2019b for further detail).

3.3. Memory saving

Unlike the fast AMM of Wood et al. (2015), our REML is not yet feasible for millions of observations because of memory consumption by the basis matrices $\{\mathbf{A}_1, \cdots, \mathbf{A}_P\}$. Wood et al. (2015) developed a QR-decomposition-based block-wise procedure for memory saving. Fortunately, the following similar simple procedure is available for our REML:

(i) The $N$ samples are divided into equally sized $H$ sub-samples.

(ii) The inner product matrices $\{m_{y,y}, \mathbf{m}_0, \mathbf{m}_{1:P}, \mathbf{M}_{0,0}, \mathbf{M}_{0,(1:P)}, \mathbf{M}_{(1:P),(1:P)}\}$ are initialized with matrices of zeros where $\mathbf{m}_{1:P} \in \{\mathbf{m}_1, \cdots, \mathbf{m}_P\}$, $\mathbf{M}_{0,(1:P)} \in \{\mathbf{M}_{0,1}, \cdots, \mathbf{M}_{0,P}\}$, and $\mathbf{M}_{(1:P),(1:P)} \in \{\mathbf{M}_{1,1}, \cdots, \mathbf{M}_{1,P}, \mathbf{M}_{2,1}, \cdots, \mathbf{M}_{2,P}, \cdots, \mathbf{M}_{P,1}, \cdots, \mathbf{M}_{P,P}\}$.

(iii) The following calculation is repeated for each sub-sample $h \in \{1, \cdots, H\}$:

   (iii-1) Basis matrices $\{\mathbf{A}_1^{(h)}, \cdots, \mathbf{A}_P^{(h)}\}$ are calculated,

   (iii-2) The inner product matrices $\{m_{y,y}^{(h)}, \mathbf{m}_0^{(h)}, \mathbf{M}_{0,0}^{(h)}, \mathbf{M}_{0,(1:P)}^{(h)}, \mathbf{M}_{(1:P),(1:P)}^{(h)}\}$ are evaluated using $\{\mathbf{A}_1^{(h)}, \cdots, \mathbf{A}_P^{(h)}\}$ and the $h$-th sub-samples,

   (iii-3) The inner product matrices $\{m_{y,y}, \mathbf{m}_0, \mathbf{m}_{1:P}, \mathbf{M}_{0,0}, \mathbf{M}_{0,(1:P)}, \mathbf{M}_{(1:P),(1:P)}\}$ are updated by adding $\{m_{y,y}^{(h)}, \mathbf{m}_0^{(h)}, \mathbf{M}_{0,0}^{(h)}, \mathbf{M}_{0,(1:P)}^{(h)}, \mathbf{M}_{(1:P),(1:P)}^{(h)}\}$.

   (ii-4) $\{\mathbf{A}_1^{(h)}, \cdots, \mathbf{A}_P^{(h)}\}$ and $\{m_{y,y}^{(h)}, \mathbf{m}_0^{(h)}, \mathbf{M}_{0,0}^{(h)}, \mathbf{M}_{0,(1:P)}^{(h)}, \mathbf{M}_{(1:P),(1:P)}^{(h)}\}$ are discarded

(iv) The restricted log-likelihood Eq. (12) with Eqs. (16) – (18) is constructed using the evaluated inner product matrices.

(v) Variance parameters $\{\boldsymbol{\theta}_1, \cdots \boldsymbol{\theta}_P\}$ are estimated sequentially.

The largest elements that are stored during the process are $\{\mathbf{A}_1^{(h)}, \cdots, \mathbf{A}_P^{(h)}\}$, for which the dimension of $\mathbf{A}_p^{(h)}$ equals $N/H \times L_p$. The dimension is reduced by increasing $H$. In other words, our approach is scalable in terms of memory usage. Step (iii) is easily parallelized because this step iterates independent calculations for each *h*. This procedure is useful for not only memory saving but also fast parallelization. Note that because of the basis function evaluation by sub-samples, we need to use the approximate Moran eigenvectors, which can be evaluated for sub-samples (see Section 2.3.3).

Because we do not store $\{\mathbf{A}_1^{(h)}, \cdots, \mathbf{A}_P^{(h)}\}$ for memory saving, the following processing is needed after REML estimation, to restore the effects estimates:

(vi)    The following calculation is repeated for each sub-sample $h \in \{1, \cdots, H\}$:

   (vi-1) Basis matrices $\{\mathbf{A}_1^{(h)}, \cdots, \mathbf{A}_P^{(h)}\}$ are calculated

   (vi-2) Th *p*-th effects estimates for the *h*-th sub-sample is recovered by $\hat{f}(\mathbf{z}_p) = \mathbf{A}_p \mathbf{V}(\widehat{\boldsymbol{\theta}}_p) \hat{\mathbf{u}}_p$.

This step is readily parallelized, too. Figure 1 summarizes our estimation procedure, which is explained in Section 3.2 and 3.3. Note that the computational time evaluated in the next section includes the time for this step.

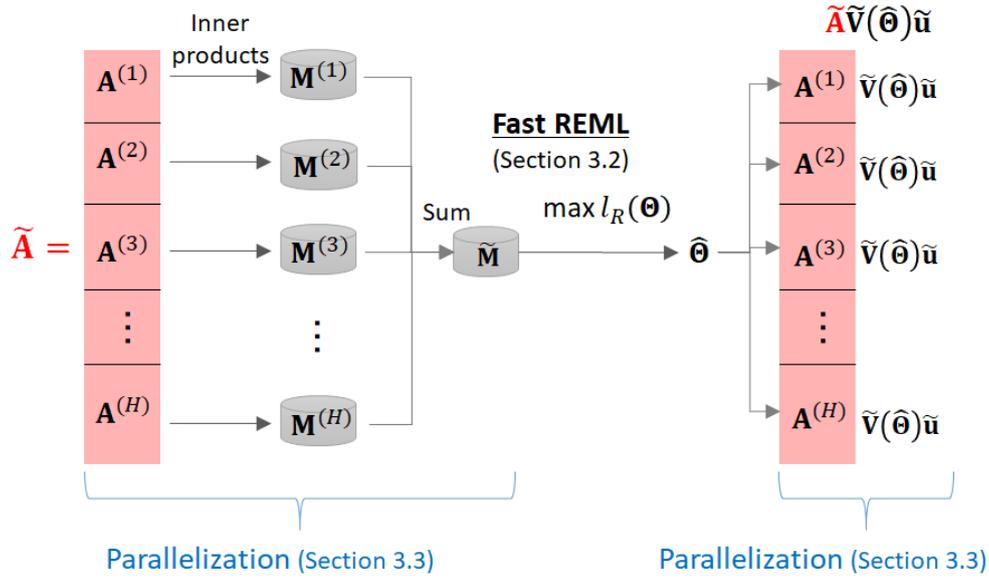

Figure 1: Overview of our fast estimation procedure. Red represents matrices whose size depends on $N$. $\mathbf{A}^{(h)} \in \{\mathbf{A}^{(h)}_1, \cdots, \mathbf{A}^{(h)}_P\}$, $\mathbf{M}^{(h)} \in \{m^{(h)}_{y,y}, \mathbf{m}^{(h)}_0, \mathbf{M}^{(h)}_{0,0}, \mathbf{M}^{(h)}_{0,(1:P)}, \mathbf{M}^{(h)}_{(1:P),(1:P)}\}$, $\widetilde{\mathbf{M}} = \sum_{h=1}^{H} \mathbf{M}^{(h)}$ e $\widetilde{\mathbf{V}}(\mathbf{\Theta})$ is a block diagonal matrix with the $p$-th block being $\widetilde{\mathbf{V}}(\boldsymbol{\theta}_p)$, and $\widehat{\widetilde{\mathbf{u}}} = [\widehat{\mathbf{u}}_1, \cdots, \widehat{\mathbf{u}}_P]'$. As shown in this figure, a large matrix $\widetilde{\mathbf{A}}$ is replace with inner products through a parallel computation. The inner products are evaluated to estimate $\mathbf{\Theta}$. Then, spatial and non-spatial effects are recovered by another parallel computation.

To summarize, this study has developed a spatial AMM, which estimates the scale of spatial processes based on the MC, and developed a REML procedure whose memory consumption is highly scalable, with the computational complexity to estimate the variance parameters being independent of $N$. This approach will be useful for large-scale spatial regression analysis. The next section compares the estimation accuracy and computational efficiency of our approach with the fast AMM of Wood et al. (2015), which is among the fastest algorithms to estimate AMM.

## 4. Spatial prediction

In Section 2.3.3, we approximate the spatial process $\mathbf{w}$ on $N$ sample sites (Eq.5) by expanding a process defined on $L$ that includes points via the Nystrom approximation. In the same manner, the spatial process at an unobserved $s_*$ is modeled as follows:

$$w(s_*) = \left(\frac{\tau^2}{\sigma^2}\right) \hat{\mathbf{e}}(s_*) \hat{\mathbf{\Lambda}}^\alpha \mathbf{u}, \qquad \mathbf{u} \sim N(\mathbf{0}_L, \sigma^2 \mathbf{I}_L). \tag{19}$$

$\hat{\mathbf{e}}(s_*)$ is derived using the Nystrom extension as

$$\hat{\mathbf{e}}(s_*) = \left[\mathbf{c}(s_*)' - \mathbf{1} \otimes \left(\frac{\mathbf{1}'_L \mathbf{C}_L}{L}\right)\right] \mathbf{E}_L (\mathbf{\Lambda}_L + \mathbf{I}_L)^{-1}, \tag{20}$$

where $\mathbf{c}(s_*)$ is a $L \times 1$ vector whose $l$-th element equals a given kernel $c(d_{*,l})$ (e.g., $\exp(-d_{*,l}/r)$). This spatial process at an arbitrary site can be modeled using Eqs. (19) and (20). The eigenvectors $[\hat{\mathbf{E}}', \hat{\mathbf{e}}'(s_*)]'$ are interpretable as a subset of spatial eigenfunctions defined on a continuous space.

Eq.(19) is useful for spatial production. For example, in the case of the SVC model, the unknown response variable $y(s_*)$ is defined as follows:

$$y(s_*) = \sum_{p=1}^{P} x_p(s_*) \left(b_p + w_p(s_*)\right) + \varepsilon(s_*), \qquad \varepsilon(s_*) \sim N(0, \sigma^2). \tag{21}$$

$x_p(s_*)$ is the $p$-th explanatory variable at the site. $y(s_*)$ is predicted with the expectation

$$\hat{y}(s_*) = \sum_{p=1}^{P} x_p(s_*)\left(\hat{b}_p + \hat{w}_p(s_*)\right), \tag{22}$$

where $\hat{w}_p(s_*) = \left(\frac{\hat{\tau}_p^2}{\hat{\sigma}_p^2}\right)\hat{\mathbf{e}}(s_*)\hat{\mathbf{\Lambda}}^{\hat{\alpha}_p}\hat{\mathbf{u}}$. Thus, our AMM can be used for spatial prediction.

## 5. Monte Carlo simulation experiment

This section compares our developed approach with alternative approaches through Monte Carlo simulation experiments. After describing assumptions, Section 4.1 presents estimation accuracy comparisons, and Section 4.2 presents computational efficiency comparisons.

Based on the discussions in Section 1, we focus on an AMM considering spatial process underlying residuals and regression coefficients, and group effects, whose absence can have a severe impact. Specifically, we examine estimation accuracy and computational efficiency by fitting our model to the synthetic data generated from:

$$\mathbf{y} = \sum_{p=1}^{6} \mathbf{x}_p \circ \mathbf{w}_p + \mathbf{g} + \mathbf{w}_0 + \boldsymbol{\varepsilon} \qquad \mathbf{g} \sim N(\mathbf{0}, \tau_g^2 \mathbf{I}_g) \qquad \boldsymbol{\varepsilon} \sim N(\mathbf{0}, \sigma^2 \mathbf{I}),$$

$$\mathbf{w}_0 = \hat{\mathbf{E}}_+ \boldsymbol{\gamma}_0 \qquad \boldsymbol{\gamma}_0 \sim N(\mathbf{0}, \tau^2 \hat{\mathbf{\Lambda}}_+), \tag{23}$$

$$\mathbf{w}_p = \begin{cases} \mathbf{1} + \hat{\mathbf{E}}_+ \boldsymbol{\gamma}_p & \boldsymbol{\gamma}_p \sim N(\mathbf{0}, \tau^2 \hat{\mathbf{\Lambda}}_+^3) & \text{if } p \le 3 \\ \mathbf{1} + \hat{\mathbf{E}}_+ \boldsymbol{\gamma}_p & \boldsymbol{\gamma}_p \sim N(\mathbf{0}, \tau^2 \hat{\mathbf{\Lambda}}_+^{0.5}) & \text{otherwise} \end{cases},$$

where $\mathbf{w}_p$ captures spatially varying effects from explanatory covariates $\mathbf{x}_p$, $\mathbf{g}$ captures group-wise effects, and $\mathbf{w}_0$ captures residual spatial dependence. $\hat{\mathbf{E}}_+$ and $\hat{\mathbf{\Lambda}}_+$ are defined using eigen-pairs corresponding to all the positive eigenvalues. The resulting $\mathbf{w}_p$ is a low rank GP explaining positive spatial dependence (see Section 2.2). To examine if our approach accurately estimates spatial scales underlying SVCs, we assume a faster decay of the eigenvalues $\widehat{\boldsymbol{\Lambda}}_+^3$ implies large-scale spatial variations for the first three SVCs while $\widehat{\boldsymbol{\Lambda}}_+^{0.5}$ implies small-scale variations for the other SVCs. The moderate-scale variation is assumed using $\widehat{\boldsymbol{\Lambda}}_+$ for the residual spatial dependence $\mathbf{w}_0$. $\sigma = 0.3[\text{SE of } \sum_{p=1}^{6} \mathbf{x}_p \circ \mathbf{w}_p + \mathbf{g} + \mathbf{w}_0]$ is assumed for the residual standard error. Our preliminary analysis suggests that the simulation results are similar even if $\sigma^2$ is changed. For the group effects, we define $N/20$ groups that are defined by 20 randomly selected samples.

The explanatory variables are generated from

$$\mathbf{x}_p = (1 - s_x)\boldsymbol{\varepsilon}_{x(p)} + s_x v_x \widehat{\mathbf{E}}_+ \boldsymbol{\gamma}_{x(p)}, \quad \boldsymbol{\varepsilon}_{x(p)} \sim N(\mathbf{0}, \mathbf{I}), \quad \boldsymbol{\gamma}_{x(p)} \sim N(\mathbf{0}, \widehat{\boldsymbol{\Lambda}}_+). \quad (24)$$

$v_x$ is a scaler satisfying $\text{Var}\left[v_x \widehat{\mathbf{E}}_+ \widehat{\boldsymbol{\gamma}}_{x(p)}^{(iter)}\right] = 1$, where $\widehat{\boldsymbol{\gamma}}_{x(p)}^{(iter)}$ is the replicated $\boldsymbol{\gamma}_{x(p)}$ in the *iter*-th iteration, and $s_x$ represents the proportion of the spatially dependent variation. Spatial variation in $\mathbf{x}_p$s can confound with the spatial variations in $\{\mathbf{w}_0, \mathbf{w}_1, \cdots \mathbf{w}_p, \mathbf{g}\}$; such a confounding makes estimation unstable (see Hughes and Haran 2013). Estimation of $\{\mathbf{w}_0, \mathbf{w}_1, \cdots \mathbf{w}_p, \mathbf{g}\}$ is more difficult in cases with larger $s_x$.

Parameters are estimated 200 times while varying the samples size $N \in \{500, 1000, 3000, 8000, 16000\}$, the strength of the group-wise variations $\tau_g^2 \in \{0.0, 0.5\tau^2, \tau^2\}$, and the strength of the spatial dependence in $\mathbf{x}_l$s, $s_x \in \{0.0, 0.5\}$. Besides, later, we assume a sample size of up to 10 million to evaluate the computational efficiency of our approach (Section 4.3).

As far as we have examined, the bam function in the mgcv package is among the fastest R functions to estimate AMMs. Given that, this simulation compares the developed Moran eigenvector-based AMM (M-AMM) with AMM and GeoAMM, which are implemented using this function. The AMM uses the 2-dimensional tensor product smoother, which is often used for spatial modeling. The standard GeoAMM uses GPs, which are rank reduced using the radial basis function-based approach (see Section 2.1), [4] to estimate $\{\mathbf{w}_0, \mathbf{w}_1, \cdots, \mathbf{w}_P\}$. The number of basis functions in the GeoAMM is given by 30 following a suggestion of Wiesenfarth and Kneib (2010) that 20 – 40 basis functions is a suitable default choice. In addition to these standard specifications, to compare our approach with GeoAMM in a similar settings, we also consider GeoAMM*, which considers the same number of basis functions as the M-AMM. To analyze the influence of ignoring group effects, we also estimate our model without group effects (M-AMM$_{-g}$). All of these models are estimated using REML. Based on simulation results in Murakami and Griffith (2019a), up to 200 eigen-pairs corresponding to large positive eigenvalues (basis functions) are considered in M-AMM and M-AMM$_{-g}$.

We use a Mac Pro (3.5 GHz, 6-Core Intel Xeon E5 processor with 64 GB of memory). R (version 3.6.2; https://cran.r-project.org/) is used for the model estimation. The mgcv package (version 1.8.28) is used to estimate AMM and GAMM and the spmoran package (version 0.1.6.2) is used to estimate M-AMM$_{-g}$.

---

[4] We prefer the radial basis function-based GP approximation (see Kammann and Wand, 2003) because it is one of the most popular GP approximations (Liu et al., 2019). In geostatistics, the predictive process modeling (Banerjee et al., 2008) and fixed rank kriging (Cressie and Johannesson, 2008) assume radial basis functions too. Although soap film smoothing (Wood, 2008), the stochastic partial differential equation approach (Rue et al., 2009), and other approaches are also available for low rank GP modeling, they are relatively computationally demanding. Thus, we adopt the radial basis function approach for the SVC modeling in GeoAMM and GeoAMM*.

5.1. Results: Estimation accuracy

The accuracy of estimated SVCs is evaluated by the root mean squared error (RMSE), which is given by

$$RMSE(\mathbf{w}_p) = \sqrt{\frac{1}{200N} \sum_{iter=1}^{200} \sum_{i=1}^{N} (\widehat{w}_{i,p}^{(iter)} - w_{i,p})^2} \qquad (25)$$

where *iter* represents the iteration number, $w_{i,p}$ is the *i*-th element of $\mathbf{w}_p$, and $\widehat{w}_{i,p}^{(iter)}$ is the estimate given in the *iter*-th iteration.

Figure 2 compares RMSEs of the SVCs, $\mathbf{w}_p$, when explanatory variables are spatially independent ($s_x = 0.0$). AMM, which is a simple spline approach, has the largest RMSEs across cases. The result suggests that AMM is too simple to model spatial process accurately. For small-scale SVCs, M-AMM and GeoAMM* indicate smaller RMSEs than the others. RMSEs of these two approaches are quite similar. It is verified that M-AMM is a good alternative to GeoAMM*. Remember that M-AMM estimates spatial scale using the *α* parameter while GeoAMM* does not; the result also shows that the estimation accuracy for the small-scale SVCs is unchanged even if their spatial scales are estimated. By contrast, for large-scale SVCs, RMSEs of M-AMM are smaller than GeoAMM*. It is suggested that the scale estimation is important to accurately estimate large scale SVCs.

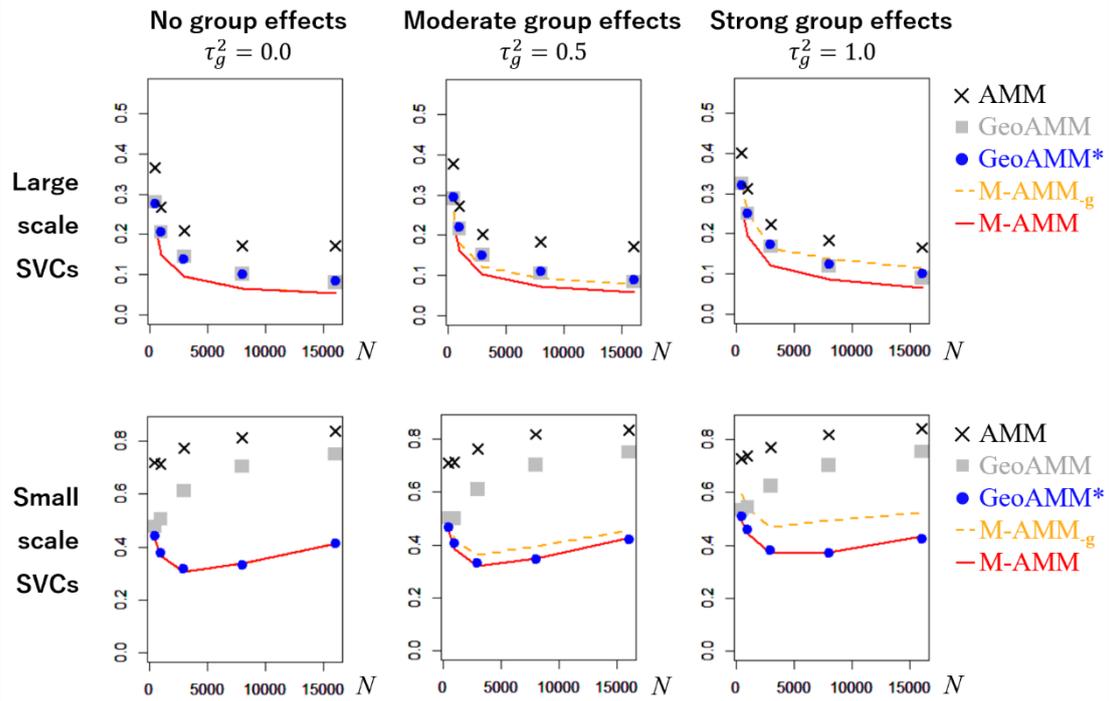

Figure 2: RMSEs for the SVCs in cases with independent explanatory variables

Figure 3 displays the RMSEs for the SVCs in cases with spatially dependent explanatory variables ($s_x = 0.5$). While the results are similar to the results when $s_x = 0.0$, the RMSEs for M-AMM$_{-g}$, which ignores the group effects, rapidly increases as the group effects increase in this case. This accuracy decrease is especially severe for the large-scale SVCs. It might be because the large-scale SVCs incorrectly estimate the ignored group effects as spatially varying effects. While discussion about group effects is quite limited in spatial statistics, especially in the context of SVC modeling, this result highlights the importance of considering group effects.

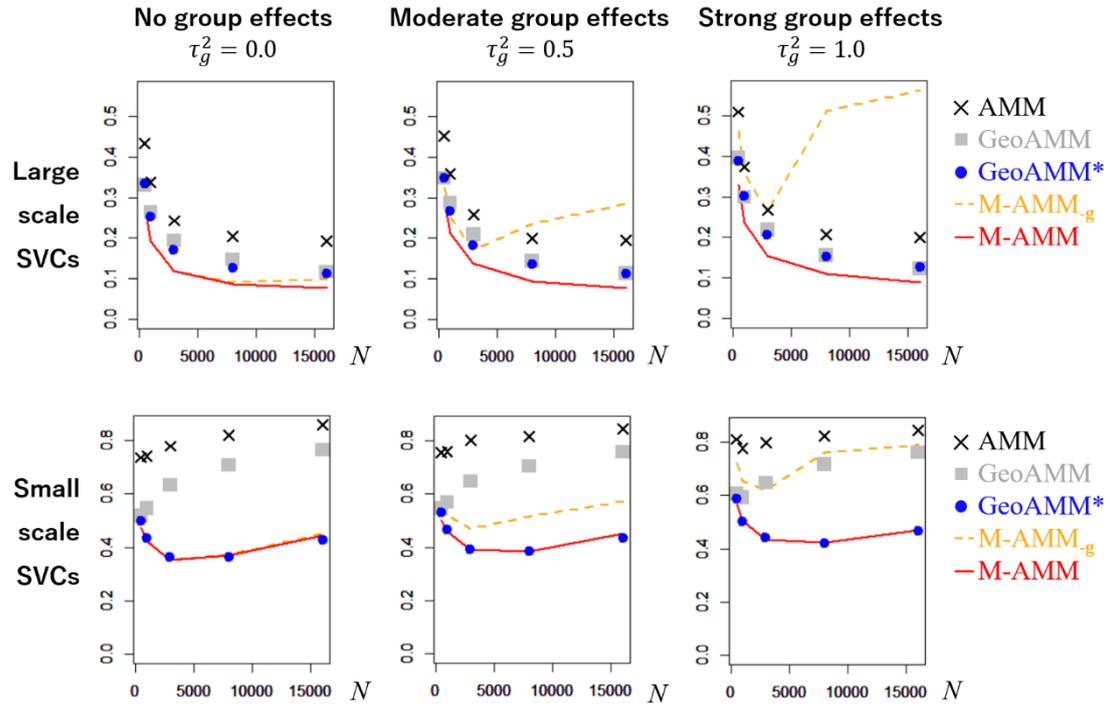

Figure 3: RMSEs for the SVCs in cases with spatially dependent explanatory variables

Figure 4 plots the first large-scale SVCs $\hat{\mathbf{w}}_1$ estimated at the first iteration when $N = 16{,}000$, $\tau_g^2 = \tau^2$, and $s_x = 0.5$. Because SVCs are usually interpreted through visual assessment, it is important to have a map pattern similar to the true $\mathbf{w}_1$. Unfortunately, the SVCs estimated by M-AMM$_{-\mathbf{g}}$ have too much small-scale variation relative to the true patterns. This scale misspecification is attributable to the absence of the group effects term. While the GeoAMM estimates are visually similar to the true patterns, the GeoAMM* estimates have somewhat noisy patterns. The noise is attributable to the absence of the scale parameter for the SVCs. By contrast, the proposed approach, which estimates the scale, accurately specifies the map pattern of the large-scale SVCs. The scale parameter is found to be helpful to estimate large-scale SVCs.

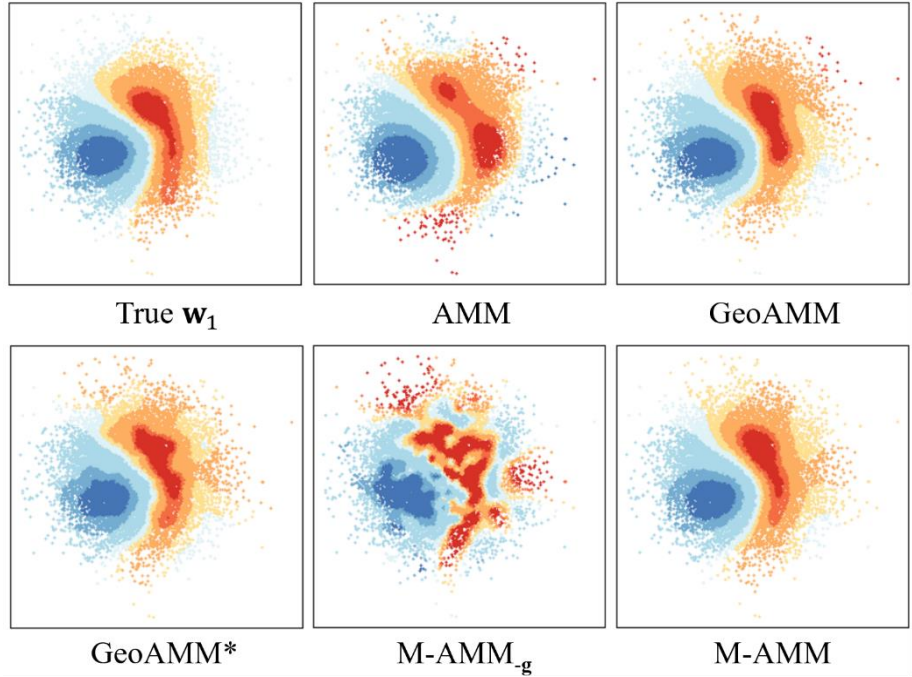

Figure 4: Large-scale SVCs estimated in the first iteration ($N$ = 16,000, $\tau_g^2 = \tau_s^2$, and $w_{xs} = 0.5$)

Figure 5 plots the first small-scale SVCs $\widehat{\mathbf{w}}_4$ estimated in the first trial. Unfortunately, AMM and GeoAMM, which are widely used, result in overly smoothed SVC estimates. Although 20 – 40 basis functions are alluded to as sufficient in AMM studies, many more basis functions might be required to avoid such an over-smoothing. GeoAMM*, M-AMM$_{-g}$, and M-AMM, which consider a larger number of basis functions, estimate the small-scale SVCs accurately. Yet, M-AMM$_{-g}$ has patterns with that partly differ from true pattern; the difference is again attributable to the absence of the group effects term.

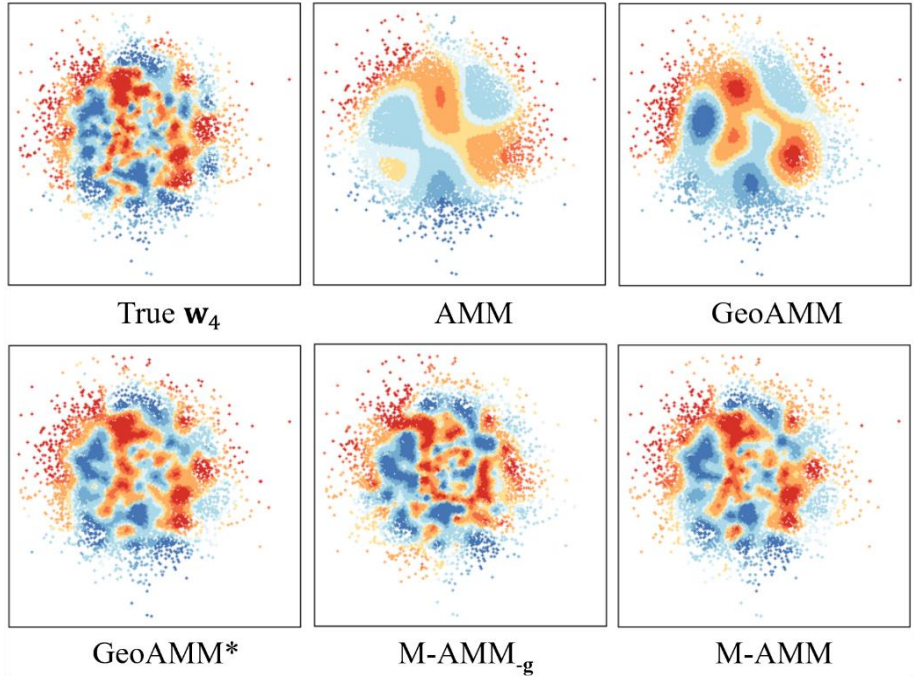

Figure 5: Small-scale SVCs estimated in the first iteration ($N$ = 16,000, $\tau_g^2 = \tau_s^2$, and $w_{xs} = 0.5$)

Although we have assumed that our model is the data generating process, we performed another Monte Carlo simulation study assuming GeoAMM* to be the data generating process. This latter generating process is identical to Eq.(23) with $\tau_g^2 = 1.0$, except that the SVCs are generated using the low rank GP (Eq.2), which is assumed in GeoAMM* (and GeoAMM). The mean RMSEs of the SVCs for M-AMM and GeoAMM* are plotted in Figure 6. Because GeoAMM* assumes the same structure/scale for the SVCs, the RMSE values are averaged across all the SVCs. This figure shows that the RMSEs for the two approaches are almost the same. This outcome confirms that our approach is as accurate as the GeoAMM*, even in this case.

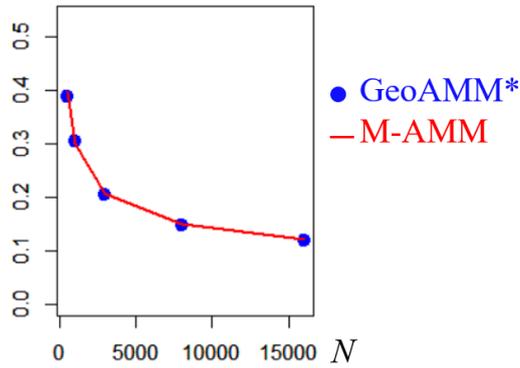

Figure 6: RMSEs for the SVCs when assuming GeoAMM* as the true data generating process

Figure 7 plots the RMSEs for the estimated group effects $\hat{\mathbf{g}}$ when $s_x = 0.5$. AMM and GeoAMM have large RMSEs, which are attributable to the mis-specification of the small-scale SVCs. By contrast, GeoAMM and M-AMM estimate $\hat{\mathbf{g}}$ fairly accurately across cases. Their accuracy is portrayed by Figure 8, which compares the true and estimated group effects obtained when $N = 16{,}000$, $\tau_g^2 = \tau^2$, and $s_x = 0.5$.

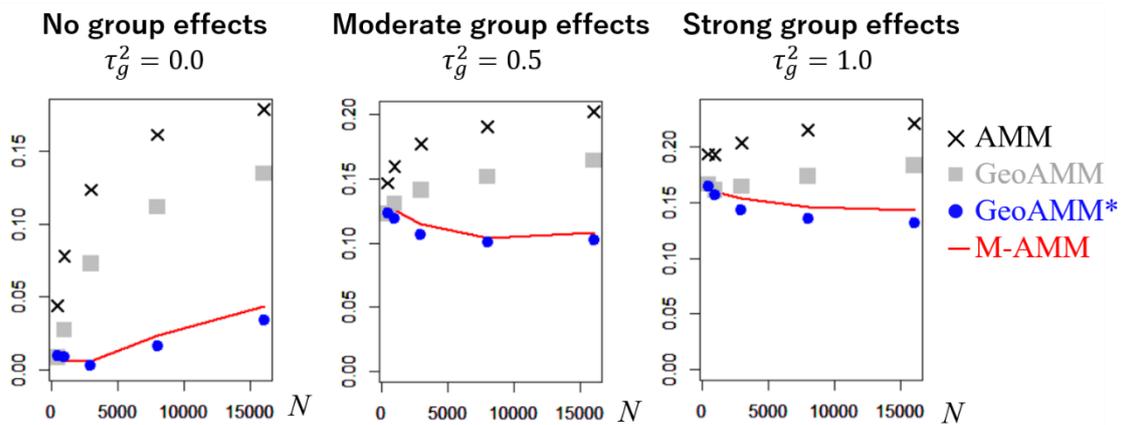

Figure 7: RMSEs for the group effects in cases with spatially dependent explanatory variables

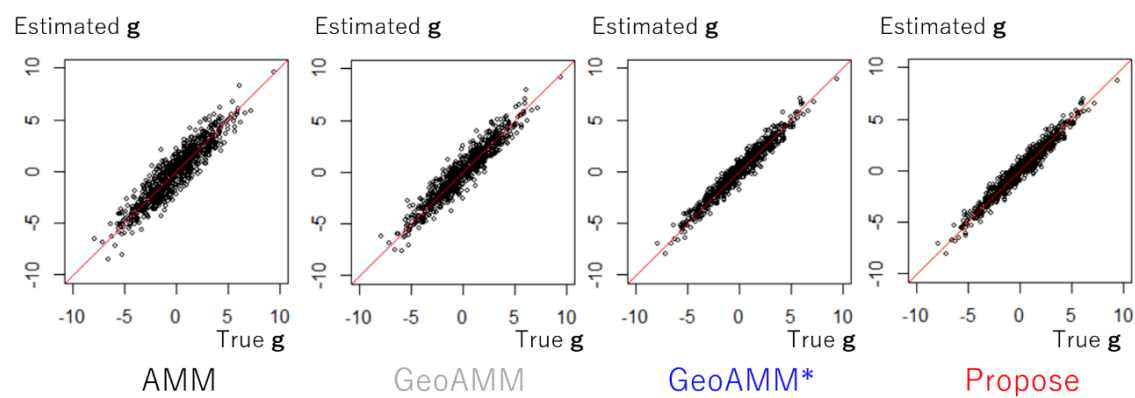

Figure 8: True and estimated group effects in the first iteration

In summary, AMM and GeoAMM, which are standard specifications, suffer from an over-smoothing of the small-scale SVCs and poor estimation for group effects. M-AMM$_{-g}$, which ignores group effects just like standard geostatistical models, suffers from a severe mis-specification of the SVCs in the presence of spatial dependence in explanatory variables. GeoAMM* and M-AMM accurately estimate both SVCs and group effects, although M-AMM outperforms GeoAMM* in terms of large-scale SVC estimation.

These results suggest that M-AMM is an attractive alternative of GeoAMM or GeoAMM*. In particular, M-AMM is more accurate than GeoAMM in the standard setting.

5.2. Results: computational time

This section compares the computational times of GeoAMM$_0$ and GeoAMM with M-AMM when $\tau_g^2 = \tau_s^2$, and $s_x = 0.5$. M-AMM is parallelized over 12 cores. In addition, we compare M-AMM without the memory efficient procedure (M-AMM$_0$), which is equivalent to the procedure of Murakami and Griffith (2019b). Because the groups, which are assumed to increase as $N$ increase, can make interpretation of the comparison results difficult, we assume no group effects (**g** = **0**) in this section. The AMM, GeoAMM, and GeoAMM* are also parallelized using these cores on the bam function in the mgcv package. We use the default setting of the bam function implementing the algorithm of Wood et al. (2015). These models are estimated 5 times in each case, with $N \in$ {10,000, 25,000, 50,000, 100,000, 250,000, 500,000, 1,000,000}.

Figure 9 summarizes the comparison results. M-AMM$_0$ is available only in cases with $N \leq$ 250,000. In addition, M-AMM$_0$ was the slowest. M-AMM$_0$ is not suitable for very large samples. GeoAMM$_0$, which considers only 30 basis functions for each SVC, is the firstest when the sample size is less than 0.5 million. However, when $N >$ 0.5 million, M-AMM is firster than GeoAMM$_0$ despite our approach using 200 basis functions for each SVC at the maximum. Compared to GeoAMM, which uses the same number of basis functions, our approach reduces the CP time by 73.0 % when $N$ equals 1.0 million, despite an additionally estimation of 7 scale parameters. More importantly, the CP time increase of our approach with respect to $N$ is smaller than the opponents. As a result, M-AMM took only 4,221 seconds even for 10 million observations (not shown in the figure). The computational efficiency of our approach is verified.

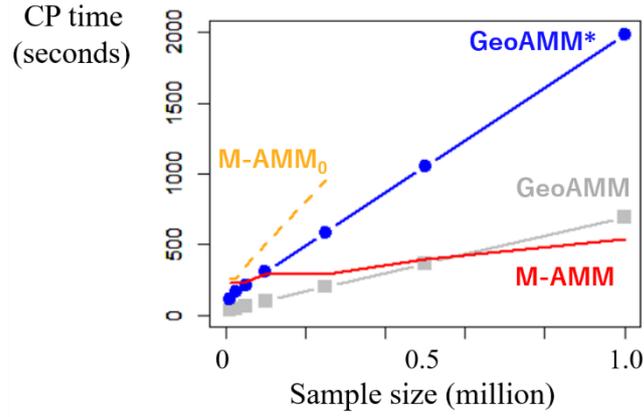

Figure 9: Comparison of computational time

## 6. Application to income analysis

### 6.1. Outline

This section applies M-AMM to a tract-level income analysis in the US. In this case, the explained variables are the household median incomes by census tract in 2015 (1,000 USA; $N = 72,160$; see Figure 10). The explanatory variables are the ratio of people over 25 who have a bachelor's degree or a higher attainment [*Univ*], the ratio of people who speak English at home [*Eng*], and the average age [*Age*], which is estimated using population by age. The SVCs for these explanatory variables are denoted by $\{\hat{\mathbf{w}}_{Univ}, \hat{\mathbf{w}}_{Eng}, \hat{\mathbf{w}}_{Age}\}$. State-level group effects and residual spatial dependences are also estimated by the M-AMM, which was used in Section 4.

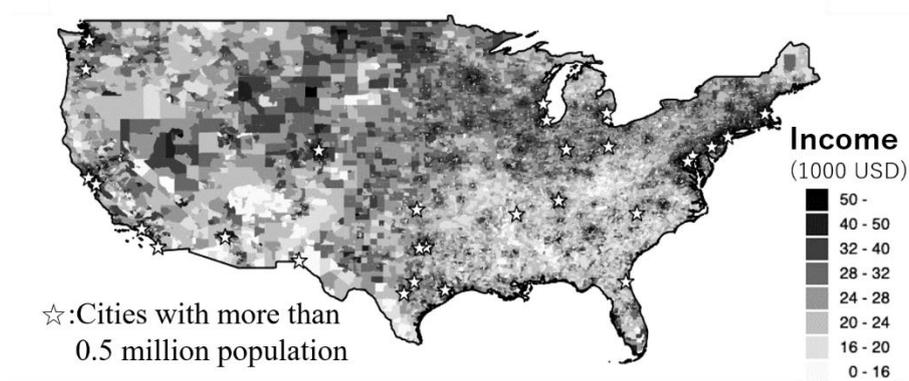

Figure 10: Tract-level household income in 2015

6.2. Result

The estimation took 103.7 seconds (including 3.62 seconds for eigen-approximation). Figure 11 plots the estimated SVCs. $\hat{\mathbf{w}}_{Univ}$ indicates large positive values near San Francisco, Los Angles, Washington D.C., New York, and other major cities. *Univ* is found to be important, especially in major cities. The $\hat{\mathbf{w}}_{Univ}$ values are also positive in other areas although they are relatively small. The estimated $\hat{\mathbf{w}}_{Eng}$ suggests that speaking English increases incomes in the northern area, while does not in the south-eastern area. People in the south-eastern area might have lower income irrespective of whether or not they speak English (see Figure 8). $\hat{\mathbf{w}}_{Age}$ suggests a positive influence of age in the north-eastern area. In this region, the influence tends to increase in non-urban areas and decrease near major cities. The importance of age in non-urban areas is estimated.

Finally, the estimated group effects are depicted in Figure 12. Based on the standard errors for the group effects, none of the states have an income-level that is significantly different from the global mean. The state-level effects are observed to be less influential relative to the spatially varying effects.

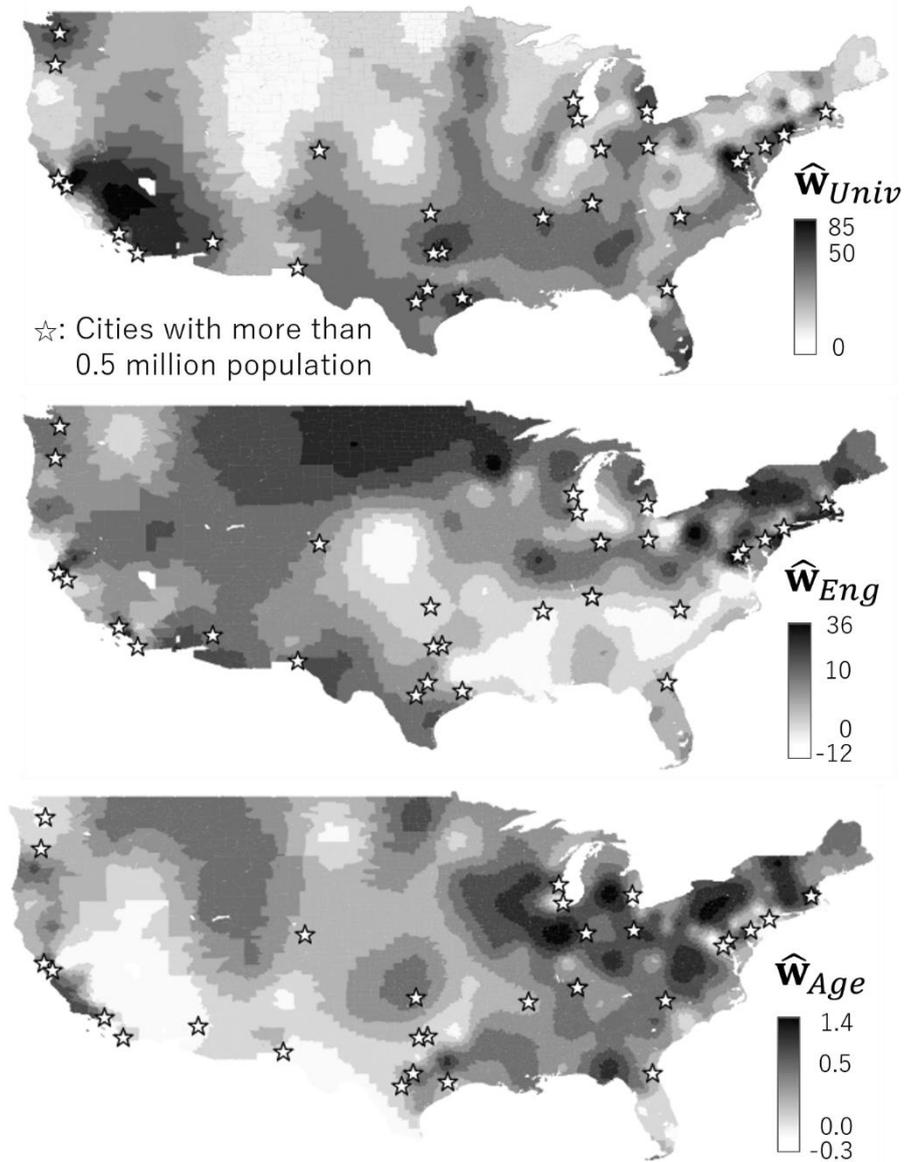

Figure 11: Estimated SVCs

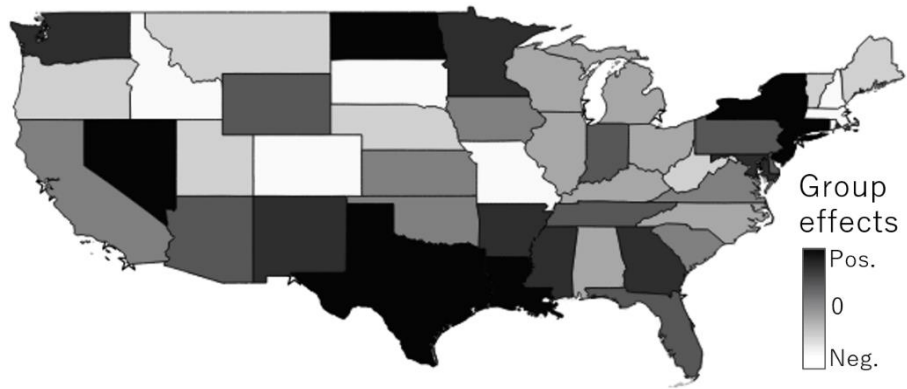

Figure 12: Estimated group effects

## 7. Concluding remarks

This paper summarizes an AMM approach for spatial modeling that is applicable to millions of observations. Unlike existing AMM, our approach explicitly estimates the scale of spatial dependence. The Monte Carlo simulations confirm the computational efficiency and estimation accuracy of our approach.

There are a wide variety of research considerations that future studies must address. First, we only considered a particular AMM with SVCs, group effects, and residual spatial dependence. Consideration of other non-spatial effects, such as temporal and smooth effects from explanatory variables is needed to verify the expandability of our approach. More effects will make identification difficult. Computationally efficient variations/effects selection, for example, using the Laplace prior, the spike and slab prior, or other priors, will be an important avenue of study in the future.

Second, we need to address the degeneracy problem of rank reduced spatial process. Stein (2014) showes that a rank reduced GP performs poorly when the true process has small-scale variations and the spatial process dominates the noise process. To mitigate this problem, multi-resolution approximation (Katzfuss 2017), the nearest-neighbor GP (Datta et al. 2016), or other scalable GP approximations might be helpful. Heaton et al. (2018) and Liu et al. (2018) review recent development of GPs for large samples.

Third, our approach must be extended to accommodate a wide variety of spatial and spatiotemporal data. An extension to non-Gaussian data will be an important first step. In fact, many data in epidemiology and ecology, where spatial data modeling is actively studied, are count data. Socioeconomic survey data are binary or ordered data in many cases. Laplace approximation might be useful to establish fast AMM for non-Gaussian data (see e.g., Wood 2017). Our approach could also be extended for spatial data fusion that combines incompatible spatial data (Gotway and Young 2002; Wang and Furrer 2019). In fact, both our model and the linear model of coregionalization (LMC; e.g., Genton and Kleiber 2015), which is a standard approach for spatial data fusion, has a linear mixed model representation. Integration of our AMM with LMC might also be a promising research topic.

R code for the Moran eigenvector-based AMM will be implemented in the R package spmoran (https://cran.r-project.org/web/packages/spmoran/index.html).

# Acknowledgement

This study is funded by the JSPS KAKENHI Grant Numbers 17K12974 and 18H03628.

# Conflict of Interest Statement

On behalf of all authors, the corresponding author states that there is no conflict of interest.